\documentclass[twocolumn]{aastex631}
\makeatletter
\let\frontmatter@title@above=\relax
\makeatother

\usepackage{amsmath}
\received{August 25, 2022}
\revised{November 11, 2022}
\accepted{November 28, 2022}

\submitjournal{ApJ}

\shorttitle{Deuterium Fractionation in TW Hya}
\shortauthors{Romero-Mirza et al.}

\graphicspath{{./}{figures/}}

\begin{document}

\title{Cold Deuterium Fractionation in the Nearest Planet-Forming Disk}

\correspondingauthor{Carlos E. Romero-Mirza}
\email{carlos.romero\_mirza@cfa.harvard.edu}

\author[0000-0001-7152-9794]{Carlos E. Romero-Mirza}
\affiliation{Center for Astrophysics $\vert$ Harvard \& Smithsonian, Cambridge, MA 02138, USA}

\author[0000-0001-8798-1347]{Karin I. Öberg}
\affiliation{Center for Astrophysics $\vert$ Harvard \& Smithsonian, Cambridge, MA 02138, USA}

\author[0000-0003-1413-1776]{Charles J. Law}
\affiliation{Center for Astrophysics $\vert$ Harvard \& Smithsonian, Cambridge, MA 02138, USA}

\author[0000-0003-1534-5186]{Richard Teague}
\affiliation{Department of Earth, Atmospheric, and Planetary Sciences, Massachusetts Institute of Technology, Cambridge, MA 02139, USA} 
\affiliation{Center for Astrophysics $\vert$ Harvard \& Smithsonian, Cambridge, MA 02138, USA}

\author{Yuri Aikawa}
\affiliation{Department of Astronomy, Graduate School of Science, University of Tokyo, Tokyo 113-0033, Japan}

\author[0000-0002-8716-0482]{Jennifer B. Bergner}
\affiliation{Department of the Geophysical Sciences, University of Chicago, Chicago, IL 60637, USA}

\author[0000-0003-1526-7587]{David J. Wilner}
\affiliation{Center for Astrophysics $\vert$ Harvard \& Smithsonian, Cambridge, MA 02138, USA}

\author[0000-0001-6947-6072]{Jane Huang}
\affiliation{Department of Astronomy, University of Michigan, 323 West Hall, 1085 S. University Avenue, Ann Arbor, MI 48109, USA}

\author[0000-0003-4784-3040]{Viviana V. Guzmán}
\affiliation{Instituto de Astrofísica, Pontificia Universidad Católica de Chile, Av. Vicuña Mackenna 4860, 7820436 Macul, Santiago, Chile}
\affiliation{N\'ucleo Milenio de Formaci\'on Planetaria (NPF), Chile}

\author[0000-0003-2076-8001]{L. Ilsedore Cleeves}
\affiliation{Astronomy Department, University of Virginia, Charlottesville, VA 22904, USA}

\begin{abstract}

Deuterium fractionation provides a window to the thermal history of volatiles in the solar system and protoplanetary disks. While evidence of active molecular deuteration has been observed towards a handful of disks, it remains unclear whether this chemistry affects the composition of forming planetesimals due to limited observational constraints on the radial and vertical distribution of deuterated molecules. To shed light on this question, we introduce new ALMA observations of DCO$^+$ and DCN $J=2-1$ at an angular resolution of $0.5"$ (30 au) and combine them with archival data of higher energy transitions towards the protoplanetary disk around TW Hya. We carry out a radial excitation analysis assuming both LTE and non-LTE to localize the physical conditions traced by DCO$^+$ and DCN emission in the disk, thus assessing deuterium fractionation efficiencies and pathways at different disk locations. We find similar disk-averaged column densities of $1.9\times10^{12}$ and $9.8\times10^{11}$ cm$^{-2}$ for DCO$^{+}$ and DCN, with typical kinetic temperatures for both molecules of 20-30K, indicating a common origin near the comet- and planet-forming midplane. The observed DCO$^+$/DCN abundance ratio, combined with recent modeling results, provide tentative evidence of a gas phase C/O enhancement within $<40$ au. Observations of DCO$^+$ and DCN in other disks, as well as HCN and HCO$^+$, will be necessary to place the trends exhibited by TW Hya in context, and fully constrain the main deuteration mechanisms in disks.

\end{abstract}


\section{Introduction} \label{sec:intro}
Isotopic fractionation studies can provide critical insights regarding the thermal history of volatile reservoirs in planet-forming environments (see \citealt{Jorgensen_2020}, \citealt{Oberg_review}, and references therein). In particular, deuterium fractionation, the enhancement of D/H in a molecule over the ISM elemental ratio of $2.3\times10^{-5}$ \citep{Linsky_2006}, has been used extensively as a chemical link between interstellar and solar system matter \citep{Ceccarelli_2014}. The high deuterium enrichment of water in Earth, for instance, has provided evidence of a cold ($<$30 K) formation pathway \citep{Hartogh_2011, Mumma_2011, Cleeves_2014}. It is still contestable, however, whether the elevated molecular D/H ratios were inherited from the pre-Solar molecular cloud, or reset by low-temperature gas-phase chemistry in the Solar Nebula. Recent sub-mm observations of DCO$^{+}$, DCN, C$_2$D, and N$_2$D$^{+}$ have revealed an active, \textit{in-situ}, deuterium fractionation chemistry in protoplanetary disks \citep[e.g.][]{Thi_2004, Qi_2008, Huang_2017, Oberg_2021, Cataldi21}. Due to poor constraints at which disk elevation the observed molecules originate, the effect of the observed disk deuterium fractionation chemistry on the composition of forming planetesimals remains unclear.

Deuterium fractionation occurs because of small energy differences between deuterated and non-deuterated molecules, which favors the formation of the deuterated isotopologue. At low temperatures ($<$30K), most deuterated molecules are expected to form via D-atom transfer reactions with H$_2$D$^+$, whereas reactions with deuterated hydrocarbons, such as CH$_2$D$^+$, are expected to drive deuterium fractionation at warmer temperatures ($<$100K) as H$_2$D$^+$ is destroyed \citep{Millar_1989, Roueff_2015}.
In circumstellar disks, if there is an active deuterium fractionation chemistry in the cold, planet-forming midplane, H$_2$D$^+$ can efficiently form other molecules, such as DCO$^+$:
\begin{equation} \label{eq:colddco}
    \mathtt{H_2D^+ + CO \rightarrow DCO^+ + H_2}
\end{equation}
\citep{Dalgarno_1984}, as long as CO is not frozen out. The deuteration can then be propagated to DCN, mainly through:
\begin{equation} \label{eq:colddcn}
\begin{split}
\mathtt{DCO^+ + HNC} & \rightarrow \mathtt{HNCD^+ + CO} \\
\mathtt{HNCD^+ + e} & \rightarrow \mathtt{DCN + H}
\end{split}
\end{equation}
\citep{Willacy_2007}. Alternatively, if the warm, CH$_2$D$^+$-driven pathway is more efficient in disks, one would expect to find DCO$^+$ in elevated molecular layers or even in the disk atmosphere, formed from gas-phase CO:
\begin{equation}
\mathtt{CH_2D^+ + CO \rightarrow DCO^+ + CH_2}
\end{equation}
\citep{Favre_2015} and DCN,
\begin{equation}
\begin{split}
\mathtt{CH_2D^+ + e} & \rightarrow \mathtt{CHD + H } \\
\mathtt{CHD + N} & \rightarrow \mathtt{DCN + H}
\end{split}
\end{equation}
\citep{Millar_1989, Turner_2001}. A detailed excitation analysis of deuterated molecules, therefore, can shed light on the relative importance of the cold and warm fractionation pathways, and hence the impact of \textit{in-situ} deuterium fractionation on the composition of planet-forming material in disk midplanes.

Most observations of DCO$^+$ in disks show extended ring-like emission structures, with the edge of the inner emission depression usually close to the edge of the millimeter dust disk \citep[e.g.][]{Qi_2008, Oberg_2015, Teague_2015, Huang_2017, Salinas_2017}, and low excitation temperatures 
($\lesssim$ 20K) \citep{Flaherty_2017}, although there is evidence that the CH$_2$D$^+$ pathway may dominate in some cases \citep{Carney_2018}. Observations of DCN in disks are more scarce, but also exhibit rings in their emission profiles, albeit less extended than DCO$^+$ \citep[][]{Huang_2017, Salinas_2017, Oberg_2021}. In particular, \citet{Cataldi21} conducted a high-resolution deuteration study of 5 Class II disks. Based on DCN/HCN ratios, \citet{Cataldi21} confirmed \textit{in-situ} deuteration via exchange reactions towards all disks, and found evidence that both cold and warm deuteration pathways are active. There are yet, however, no conclusive DCN excitation temperature measurements in disks from multi-line observations.

Located at 59.6 pc \citep{Gaia_2021}, TW Hya is on of the closest and best studied Solar Nebula analogs, and thus has been the target of multiple (marginally resolved) observations seeking to understand deuterium fractionation in disks \citep[e.g.][]{vanDishoeck_2003, Thi_2004, Qi_2008, Oberg_2012}. Recently, \citet{Oberg_2021} (henceforth Ö21) observed the $J=3-2$ and $J=4-3$ transitions of DCO$^+$ and DCN with ALMA as part of the TW Hya Rosetta Stone Project; using the new data and archival observations of the $J=2-1$ lines, Ö21 carried out a rotation diagram analysis and radiative transfer simulations, and concluded that DCO$^+$ emits from a warm, elevated molecular layer with a disk averaged excitation temperature close to 40K, which indicates that the CH$_2$D$^+$  pathway dominates throughout the disk. Additionally, Ö21 retrieved similar disk averaged column densities for DCO$^+$ and DCN of $\sim 3\times 10^{12}$ cm$^{-2}$, but could not constrain the excitation temperature of DCN due to the low S/N of the $J=2-1$ line. 

In this work, we present the first direct measurement of the kinetic temperature of DCN, and a detailed excitation analysis of both DCN and DCO$^+$ using new ALMA observations of the $J=2-1$ transitions, augmented by archival observations of the $J=3-2$ and $J=4-3$ lines. In Section \ref{sec:obs}, we describe the ALMA observations, as well as the data reduction process and imaging strategy. Section \ref{sec:exc} describes our emission line modeling and excitation analysis. Finally, a discussion of our results and a model comparison is presented in Sections \ref{sec:discussion} and \ref{sec:conclusions}.

\section{Observational Results} \label{sec:obs}

\subsection{ALMA Observations \& Calibration}
The observations of DCO${^+}$ and DCN used for this paper were carried out as part of three different ALMA programs. In particular, we present new ALMA Band 4 observations of TW Hya (PI: K. Öberg, 2019.1.01153.S) of DCO${^+}$ and DCN $J=2-1$ lines, augmented by archival Band 4 observations from project 2016.1.00440.S (PI: R. Teague) and Band 6 observations from the TW Hya as a Chemical Rosetta Stone Program (PI: L. I. Cleeves, 2016.1.00311.S and 2017.1.00769.S), spanning the $J=2-3$ and $J=3-4$ transitions of DCO${^+}$ and DCN. For details of the Band 4 and 6 archival data, we refer the reader to \citet{Oberg_2021} and \citet{Teague_2020}. 

The new 2019.1.01153.S observations were part of ALMA Cycle 7 Return to Operations (RTO) phase, and were carried out in two executions on 2021 March 23 and 26 using 38 and 39 antennas, respectively. Each execution lasted 83 minutes and integrated on source for 48 minutes at an angular resolution of 0.37". In both cases, the correlator was tuned to target DCO${^+}$ and DCN $J=2-1$ at a spectral resolution of 120 m/s, and included a wide continuum window for self-calibration spanning C$_4$H $N=14-13$, CH$_3$CN $F=8-7$, and H$_2$CS $J=4-3$ (detected), with channel widths of 250-500 m/s. For both executions, J1037-2934 and J1103-3251 were used as flux and phase calibrators, respectively.

All raw 2019.1.01153.S data were first calibrated by ALMA staff using the CASA \texttt{v6.1.15} pipeline, and the products were subsequently self-calibrated using the same version of CASA. In brief, all continuum channels with significant spectral line emission were manually flagged, and the remaining channels were averaged to create a pseudo-continuum dirty image. Each execution block was then aligned by fitting a 2D Gaussian to the dirty continuum images and merged into a single measurement set. Next, the combined continuum visibilities were deconvolved and underwent two rounds of phase-only self-calibration, using solution intervals of 720 and 360s, followed by a single round of amplitude self-calibration. Finally, the self-calibration solutions were applied to all the spectral windows containing emission lines, and the continuum emission was subtracted from each window in the \textit{uv}-plane using the CASA task \texttt{uvcontsub} with a fit order of 1.

\subsection{Imaging Strategy}

\subsubsection{CLEANing} \label{sec:CLEAN}
The synthesized and deconvolved image cubes were created in CASA, using the multi-scale CLEAN algorithm \citep{Cornwell_2008} with \texttt{scales = [0, 5, 15, 25]} pixels; as opposed to the classic \citet{Hogbom_1974} algorithm, which uses $\delta$-functions as a functional basis set, the multi-scale algorithm uses a more advanced set of $\delta$-functions and 2D circular Gaussian components as the basis functions for the CLEAN model. All DCO${^+}$ and DCN spectral lines were imaged using a Briggs parameter of 0.5 and cell size of 0.05", resulting in $4-10$ pixels per beam FWHM. The spread arises from the factor of $\sim2$ difference in the beam sizes between the $J=4-3$ and $2-1$ observations. The beam sizes, along with all relevant CLEANing variables and integrated fluxes are displayed on Table \ref{tab:clean}. In order to efficiently capture all the molecular line emission while CLEANing, the \texttt{keplerian\_mask}\footnote{github.com/richteague/keplerian\_mask} package was used to generate masks based on the projected velocity of each pixel,
\begin{equation}
    v_K = \left(\frac{GM_*r^2}{(r^2 + z^2)^{3/2}}\right)^{1/2} \sin(i)\cos(\phi),
\end{equation}
where $M_*$ is the stellar mass, assumed to be 0.81 $M_{\odot}$, and $i=5.8^o$ is  the disk inclination \citep{Teague_19}. We do not attempt to measure the emission height given the low disk inclination, and simply assume $z\ll r$ to generate the masks. The Keplerian masks were further truncated at $r=3.0"$ based on visual inspection of the channel maps.

\begin{figure*}
    \centering
    \includegraphics[width=0.99\textwidth]{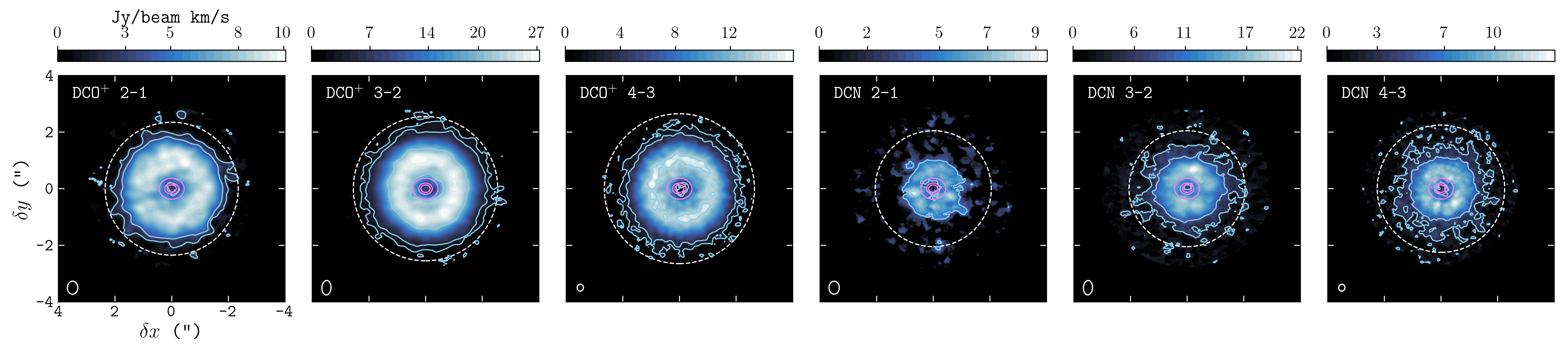}
    \caption{Full-resolution, velocity-integrated intensity maps of the $J=2-1$, $J=3-2$, and $J=4-3$ DCO$^{+}$ and DCN lines, with blue contours drawn at 5, 10, 50, 100, and 150$\times$ the mean moment zero uncertainty (see Table \ref{tab:clean}). The inner violet contours indicate the continuum emission, drawn at 50, 80, and 90$\%$ of the maximum continuum intensity of 45 mJy/beam. The outer dashed line indicates $r_{out}$ (see text) for each line, and the restored beam sizes are shown at the bottom left corner of each panel.}
    \label{fig:moment_zero}
\end{figure*}

\begin{figure*}
    \centering
    \includegraphics[width=0.99\textwidth]{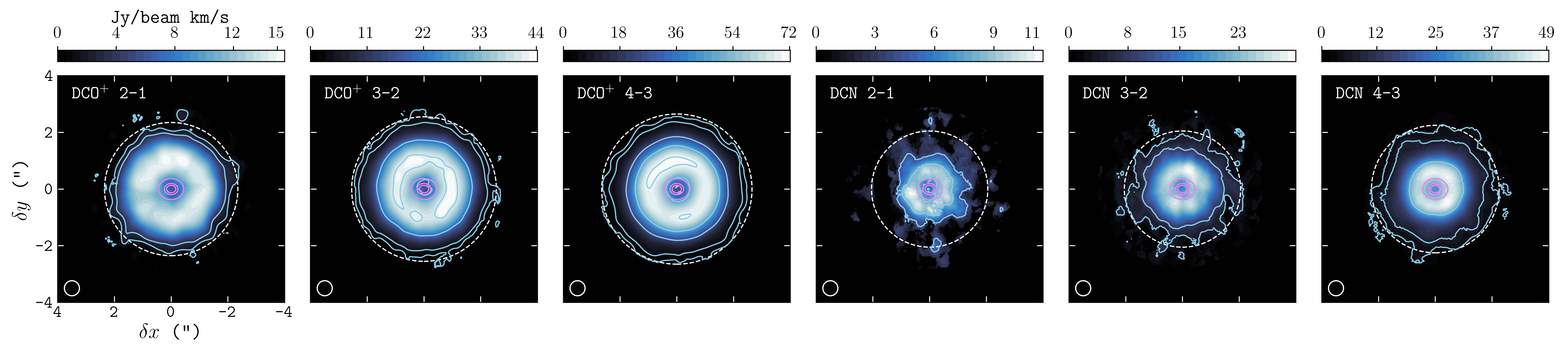}
    \caption{Same as Figure \ref{fig:moment_zero}, for the images tapered to 0.5" resolution.}
    \label{fig:moment_zero_tapered}
\end{figure*}

To deconvolve the DCO$^{+}$ visibilities, only one Keplerian mask centered on the systemic velocity was used, since all the hyperfine emission line components are blended. However, the hyperfine structure of DCN is spectrally resolved, and a set of six Keplerian masks, centered on the expected emitting frequency of each satellite line, were used to clean each $J$ transition. During the deconvolution process, the new and archival visibilities were combined and binned down to a common spectral resolution of 0.3, 0.04, and 0.07 km/s for the $J=2-1$, $J=3-2$, and $J=4-3$ lines, respectively. A primary beam correction was applied to all images after CLEANing and correcting the residual-model unit mismatch (see Section \ref{sec:JvM}). All image cubes were CLEANed to 3$\times$ the dirty rms. Finally, in order to carry out the excitation analysis described in Section \ref{sec:Modeling}, we create a second set of image cubes with tapered beams of common FWHM $= 0.5"$ by convolving the data using the CASA task \texttt{imsmooth}. Section \ref{sec:moment-zero} presents velocity-integrated emission maps and radial profiles for the full-resolution images. 

\begin{deluxetable*}{lccccc}[t!]
\tabletypesize{\footnotesize}
\tablewidth{0pt}
 \tablecaption{Deconvolution Parameters  \label{tab:clean}}
 \tablehead{
 \colhead{Transition} & \colhead{Restored Beam}  & \colhead{Dirty rms} & \colhead{Channel Width$^{\dagger}$}  & \colhead{JvM $\epsilon$} & \colhead{Flux $^\ddagger$} \vspace{-0.2cm} \\   
 \colhead{} & \colhead{(maj$\times$min", PA$^o$)}  & \colhead{(mJy/beam)} & \colhead{(km/s)} &  \colhead{} & \colhead{(Jy km/s)}
 }
 \startdata 
 DCN 2--1 & 0.46$\times$0.37, 88.8 & 0.9 & 0.30 & 0.73 & $0.14\pm0.03$ \\
 DCN 3--2 & 0.51$\times$0.32, 86.9 & 3.2 & 0.04 &  0.69 & $0.43\pm0.09$ \\
 DCN 4--3 & 0.24$\times$0.22, 80.9 & 2.3 & 0.07 & 0.50 & $0.77\pm0.11$ \\
 DCO$^+$ 2--1 & 0.46$\times$0.37, 89.3 & 1.0 & 0.30 &  0.72 & $0.42\pm0.09$ \\
 DCO$^+$ 3--2 & 0.51$\times$0.32, 86.9 & 3.4 & 0.04 &0.70 & $1.22\pm0.09$ \\
 DCO$^+$ 4--3 & 0.24$\times$0.22, 79.7 & 2.8 & 0.07 &  0.51 & $1.97\pm0.10$ \\ 
 \enddata
 \tablecomments{The intensity rms is calculated in the unmasked regions of the dirty-image cube. The JvM $\epsilon$ parameter corresponds to the ratio between the clean and the dirty beam volumes. $\dagger$Obtained after combining all archival and new observations. $\ddagger$Integrated out to each line's $r_{out}$ using the tapered images. The reported uncertainty corresponds to $3\times$ the moment zero uncertainty and does not include the estimated 10$\%$ flux calibration error.} 
\end{deluxetable*}

\subsubsection{JvM Correction} \label{sec:JvM}
We implement an additional correction to the fluxes of the final CLEANed images due to the unit mismatch between the residual map and the CLEAN model\footnote{For a comprehensive and pedagogic review of the Jorsater \& van Moorsel correction, see \citet{Czekala_2021}.}. Originally described by \citet{Jorsater_1995}, the \textit{"JvM"} correction accounts for the fact that the residual map has units of Jy/(dirty beam), whereas the CLEAN model has units of Jy/(CLEAN beam); the divergence can be significant, even for moderately non-Gaussian dirty beams with small shelves. For our observations, the correction is made by estimating the ratio of the CLEAN and dirty beam volumes, $\epsilon$, at the location of the first null value. Although typically assumed to be significant only for observations carried out with a combination of very long and short baselines, we find significant $\epsilon$ corrections. As discussed in Section \ref{sec:discussion}, we find that applying this correction greatly improves the signal-to-noise ratio of the data products but, fortunately, does not have a critical impact in the excitation analysis.

\subsubsection{Velocity-Integrated Intensity Maps and Profiles} \label{sec:moment-zero}

We create velocity-integrated intensity (moment-zero) maps and radial profiles using the full-resolution and tapered images of DCN and DCO$^{+}$. As described in Section \ref{sec:Modeling}, these are not used to carry out our excitation analysis, but simply to analyze the emission morphology. We create the moment-zero maps by fitting a set of Gaussian line profiles to each pixel with a peak intensity greater than 3$\times$ the channel rms. To maximize the recovered emission, we fit models consisting of the sum of six Gaussians of equal line width initialized near the expected velocity of the six strongest hyperfine components of DCN and DCO$+$. While all six hyperfine lines in each $J$ level are fit simultaneously, each $J=2-1$, $J=3-2$, and $J=4-3$ profile is fit independently. The total integrated emission is then averaged in 0.1" ($\sim 6$ au) radial bins to create the radial profiles, resulting in $5$ samples per beam.

To aid the interpretation of the emission profiles and ensure the cubes are well-aligned prior to our excitation analysis, we fit a 2D ring model to each tapered moment-zero image using \texttt{astropy}'s simplex least-squares fitter to obtain small emission center corrections, $\Delta x$ and $\Delta y$, and an inner ring radius $r_{in}$. Instead of accepting the outer ring radii returned by the simplex fitter, which we find tend to underestimate the full emission extent, we estimate the outer ring radius based on the integrated intensity uncertainty 
\begin{equation}
\sigma_{M0} = \texttt{rms}\times \sqrt{n_{chan}} \times \Delta v,
\end{equation}
where $n_{chan}$ is the number of channels in which each pixel lies inside the Keplerian mask and $\Delta v$ is the channel width.  In particular, the outer ring radius for each image is estimated as the first radial bin in which the integrated intensity drops below $3\sigma_{M0}$. Table \ref{tab:ringmodels} lists the best fitting ring parameters for each image.


\begin{deluxetable}{ccccc}[t!]
\tabletypesize{\footnotesize}
\tablewidth{0pt}

 \tablecaption{Best Fit 2D Ring Models \label{tab:ringmodels}}

 \tablehead{
 \colhead{Transition} & \colhead{$\Delta x$ (")}  & \colhead{$\Delta y$ (")} & \colhead{r$_{in}$ (")} & \colhead{r$_{out}$ (")} 
 }

 \startdata 
 DCN 2--1 & -0.05 & 0.01 & 0.22 & 2.05 \\
 DCN 3--2 & 0.00 & -0.07 & 0.26 & 2.05 \\
 DCN 4--3 & -0.02 & -0.02 & 0.25 & 2.25 \\
 DCO$^+$ 2--1 & -0.02 & 0.00 & 0.34 & 2.35 \\
 DCO$^+$ 3--2 & 0.00 & -0.08 & 0.31 & 2.55 \\
 DCO$^+$ 4--3 & 0.00 & -0.03 & 0.28 & 2.65 \\
 \enddata

 \vspace{-0.5cm}
 \tablecomments{}
\end{deluxetable}

\begin{figure}[t!]
    \centering
    \includegraphics[width=0.45\textwidth]{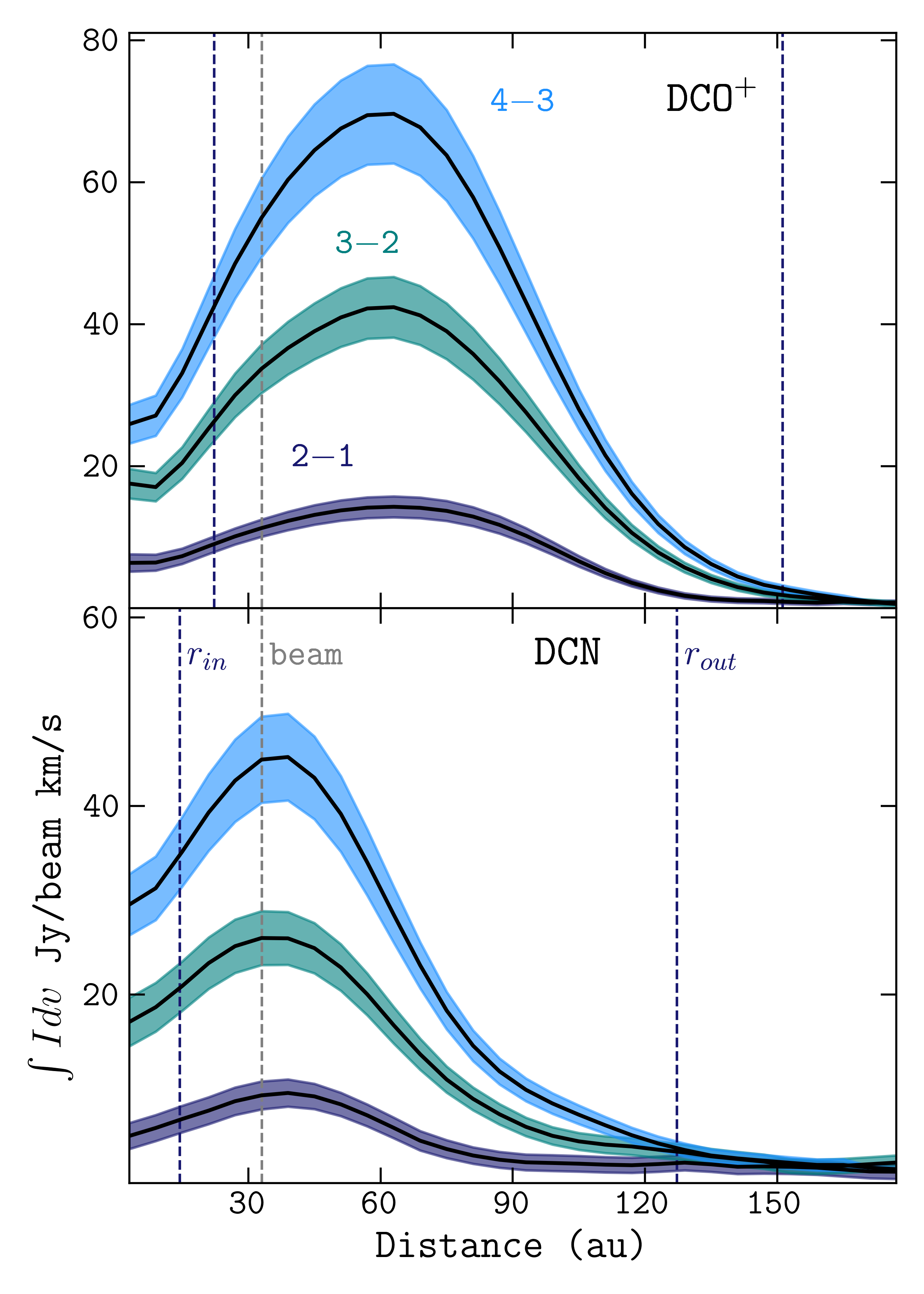}
    \caption{\texttt{Top:} Velocity-integrated flux density profiles for the $J=2-1$, $J=3-2$, and $J=4-3$ DCO$^+$ lines, using the tapered data products. Shaded regions indicate the 1$\sigma$ rms-based uncertainty with the estimated 10\% flux calibration error added in quadrature. The dotted gray line indicates circularized beam FWHM, and the inner and outer ring radii for each species, averaged for all $J$-transitions, are shown as dashed blue lines. \texttt{Bottom:} same as the top panel, for DCN.}
    \label{fig:intprofiles}
\end{figure}

Figure \ref{fig:moment_zero} shows the full-resolution moment-zero maps of each observed DCO$^+$ and DCN transition, while Figure \ref{fig:moment_zero_tapered} the maps made with the tapered image cubes. The dust disk emission is shown as violet contours, which distinctly overlap with the central emission depression exhibited by DCO$^+$ and DCN. Whether this overlap is due to continuum opacity or a chemical process is explored in Section \ref{sec:discussion}. Figure \ref{fig:intprofiles} shows the integrated intensity profiles for each transition using the images with tapered beam sizes of 0.5". Note that the uncertainty of the profiles includes the estimated 10$\%$ flux-calibration error. As discussed in Ö21 for the $J=3-2$ and $J=4-3$ lines, DCO$^+$ shows a clear inner plateau, spanning from approximately 50 to 70 au, and extended ring-like structure out to 160 au, while the DCN emission is more compact and exhibits a narrow ring which peaks at 40 au and a diffuse halo, with the emission dropping sharply out to 80 au, followed by a more gradual decrease out to $\sim 140$ au. Our new observations of the $J=2-1$ lines exhibit similar axisymmetric emission substructures to those of the higher energy transitions, which reinforces the idea that these substructures trace primarily column density variations, in addition to possible changes in the excitation conditions. 

\section{Excitation Analysis} \label{sec:exc}

\subsection{Emission Modeling} \label{sec:Modeling}

\begin{figure*}[t!]
    \centering
    \plotone{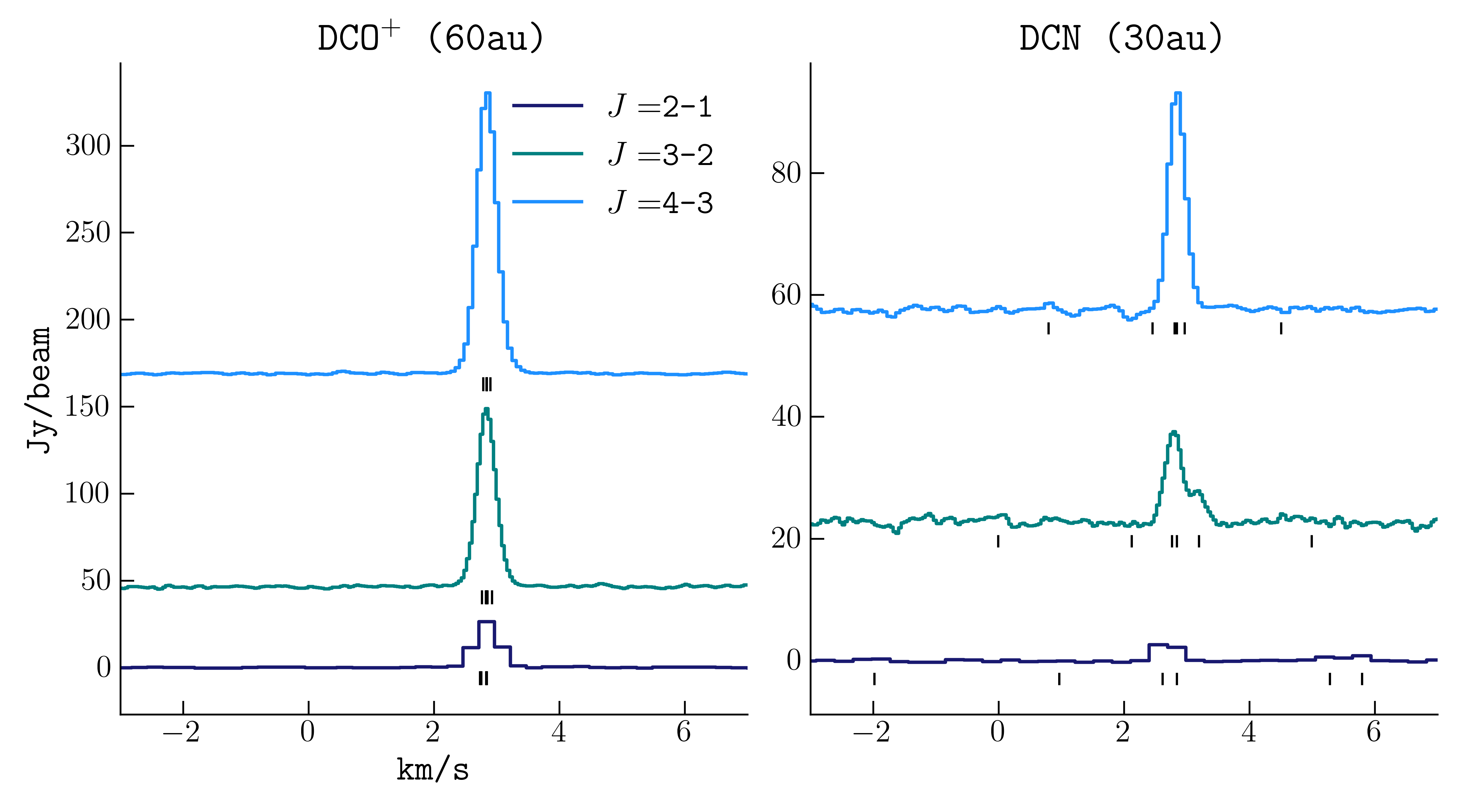}
    \caption{Stacked and deprojected spectra of all three DCO$^{+}$ (left) and DCN (right) transitions, extracted from the radial bin in which the emission of each molecule peaks (60 and 30 au, respectively). Each hyperfine structure component is indicated with a black line. The spectra are vertically shifted to be shown in the same plot.}
    \label{fig:60auspec}
\end{figure*}

Rather than relying only on the integrated intensity of the lines to estimate abundances and excitation temperatures---which was the approach used by Ö21---we attempt to maximize the information content of our observations by forward modeling the spectra itself. To this end, we use \texttt{GoFish} \citep{GoFish} to extract stacked and deprojected spectra from azimuthal bins of width 0.1", corresponding to 5 samples per beam FWHM, out to 2.5" for DCN and 3.0" for DCO$^{+}$, based on the values of $r_{out}$ reported in the previous Section. In particular, each pixel coordinate $(x,y)$ is deprojected onto a disk coordinate $(\theta, r)$ to create the bins, and the spectra extracted from these are then shifted to a common center based on our Keplerian velocity model and finally stacked. Figure \ref{fig:60auspec} shows example spectra for each transition, extracted from the 0.95-0.15" radial bin. While the DCN spectra exhibits clear satellite lines and significant asymmetric broadening, all the hyperfine structure of DCO$^{+}$ is unresolved. Nonetheless, we find it necessary to include all six hyperfine components with available spectroscopic data of each DCO$^{+}$ $J$--transition in our modeling in order to properly recover the observed line profiles. 

The line modeling methodology is described in detail in Appendix \ref{sec:modelapx}. In brief, the total emission for each $J-$transition is modeled as a sum of thermal-broadened Gaussian profiles, each of which is allowed to become saturated at the line center to account for moderate optical depths. To explore the significance of non-LTE effects, the level populations to generate the spectra are calculated with \texttt{RADEX} \citep{vandertak_2007}. Using the nested sampling package \texttt{dynesty} \citep{Speagle_2020}, we estimate posterior distributions for the gas kinetic temperature $T_{kin}$, column density $N_{T}$, the H$_2$ number density $n(H_2)$, and the systemic velocity $v_{sys}$. Finally, to account for any non-physical (primarily due to beam-smearing) line broadening sources, the modeled spectra are convolved with a Gaussian of width $\delta$, such that the total flux is redistributed to match the observed line widths. The width $\delta$ is treated as an additional free parameter. We test the excitation conditions inferred with RADEX by repeating the analysis in LTE ($T_{kin} = T_{ex}$), and obtain almost identical results. Ultimately, we choose to report the non-LTE results as they provide more conservative uncertainties and minimize the assumptions in our model.

\subsection{Spectral Fitting Results} \label{sec:fitres}

\begin{figure}[ht!]
    \centering
    \plotone{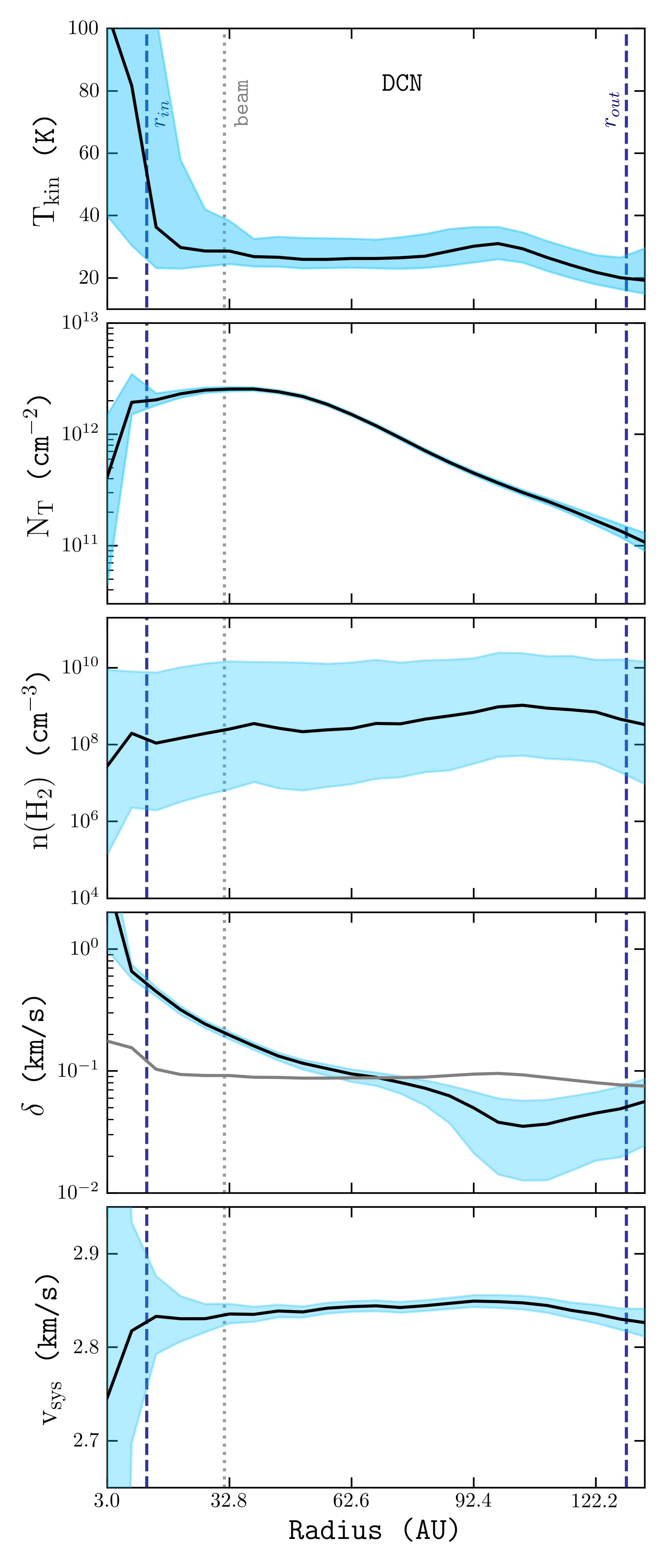}
    \caption{Radially resolved excitation conditions of DCN. \texttt{From top to bottom:} gas kinetic temperature, column density, local H$_2$ number density,  width of the Gaussian kernel applied to match the observed line widths, and systemic velocity of the stacked spectra. The shaded regions enclose the 0.16 and 0.84 posterior quantiles, while the blue dashed lines indicate the mean inner and outer ring radii. The beam FWHM is shown as a dotted line. The gray line on panel four indicates the thermal line width corresponding to the best fit kinetic temperature in the top panel.}
    \label{fig:DCN_profiles}
\end{figure}

\begin{figure}[t!]
    \centering
    \plotone{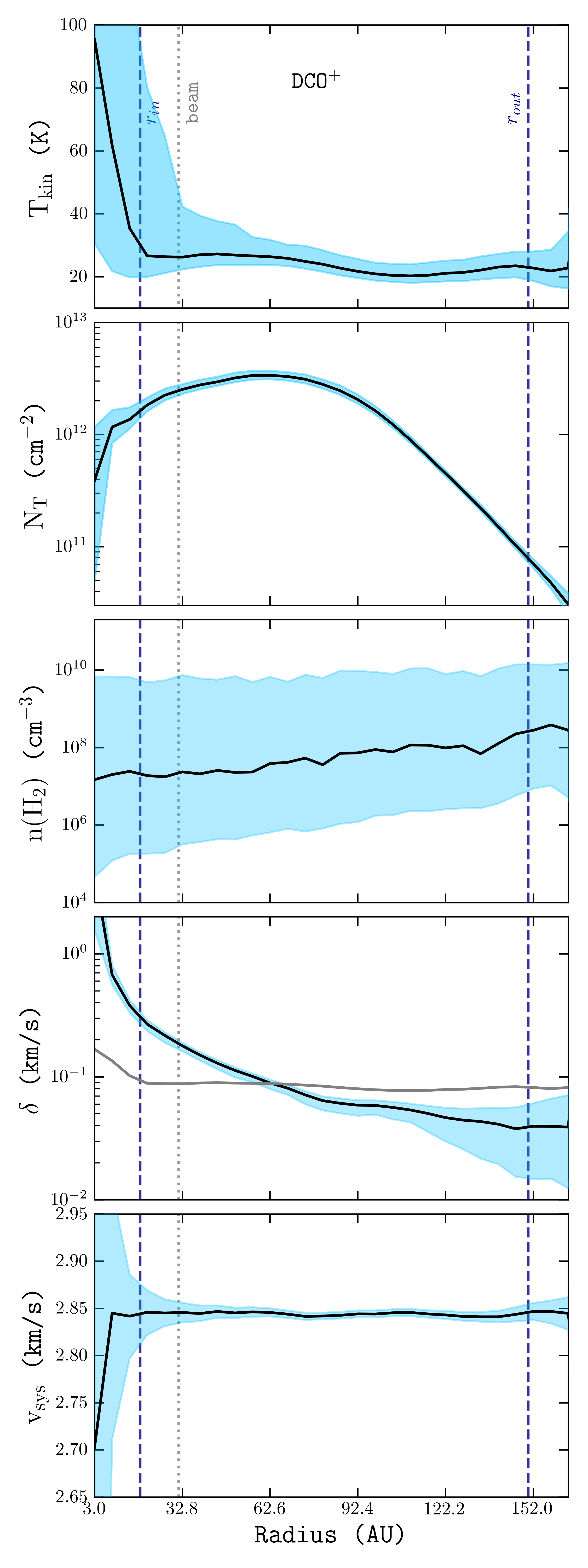}
    \caption{Same as Figure \ref{fig:DCN_profiles}, for DCO$^{+}$.}
    \label{fig:DCO+_profiles}
\end{figure}

The derived column density, kinetic temperature, H$_2$ density, line-center velocity and smoothing kernel radial profiles are shown in Figures \ref{fig:DCN_profiles} and \ref{fig:DCO+_profiles} for DCN and DCO$^+$, respectively. Two vertical dashed lines are shown in each panel of both figures, indicating average $r_{in}$ and $r_{out}$ fit to each molecule, together with the circularized beam FWHM. The radial profiles---in particular for $T_{kin}$---derived outside the region enclosed by $r_{in}$ and $r_{out}$ are likely not well constrained, since the emission is evidently unresolved at small radii and the S/N simply is too poor beyond $r_{out}$. A mechanistic explanation for the inner and outer radii exhibited by the intensity and density profiles is discussed in Section \ref{sec:discussion}. As shown in the bottom panel of Figures \ref{fig:DCN_profiles} and \ref{fig:DCO+_profiles}, the beam smearing effects are too severe to accurately fit $v_{sys}$ to spectra extracted at small radii. In the region interior to the beam width, the contributions of $\delta$ to the line widths become quite significant, which makes the temperature and gas density difficult to estimate. However, we are able to obtain reliable excitation fits for most of the disk, in a region spanning $\sim 90$ au for DCN and $\sim 120$ au for DCO$^+$. Figures \ref{fig:dcngal} and \ref{fig:dcopgal} in Appendix \ref{sec:apb} show a gallery of azimuthally averaged spectra and their best fit models.

\begin{figure}[ht!]
    \centering
    \includegraphics[width=0.45\textwidth]{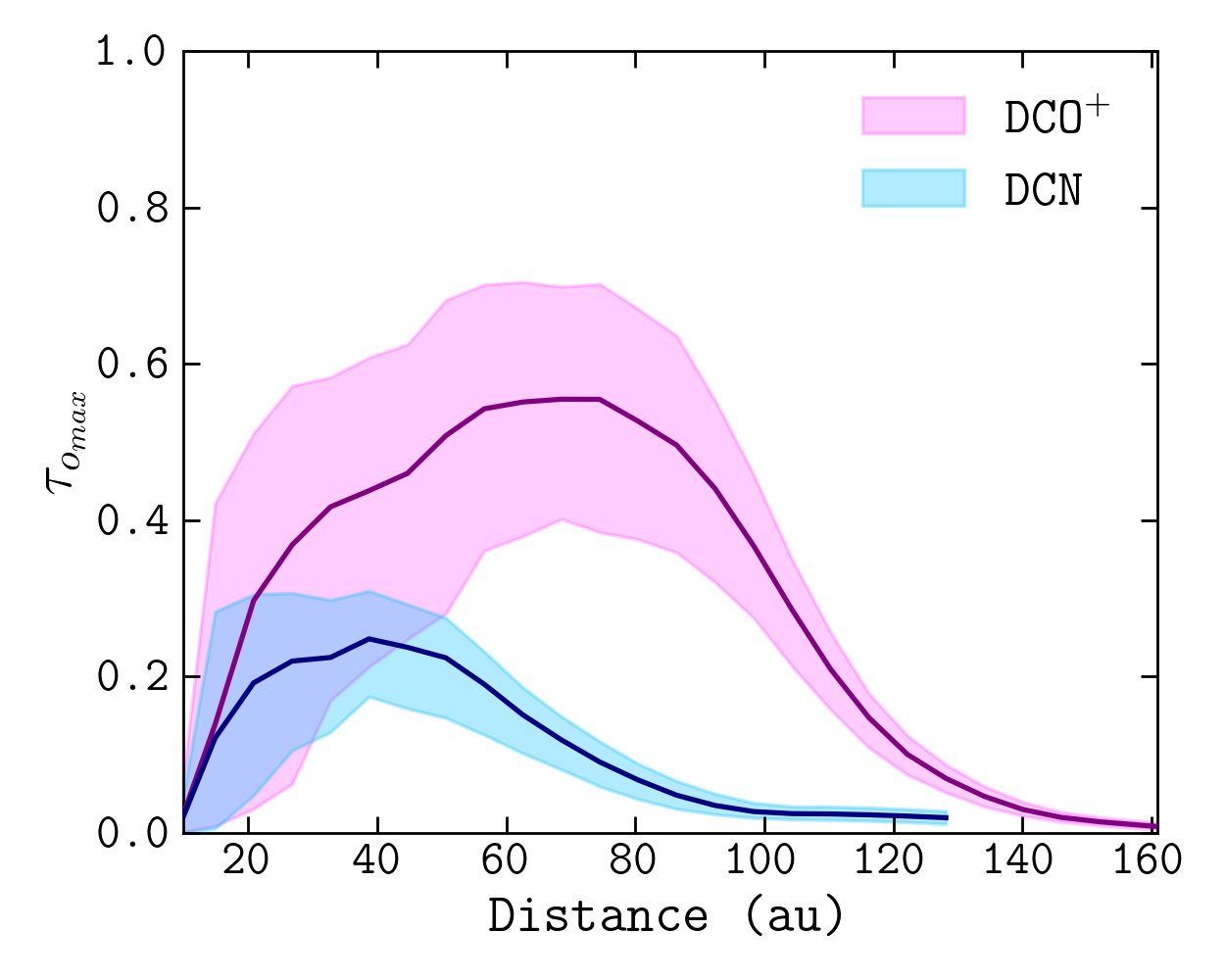}
    \caption{Radial profiles of the line-center optical depth for the most optically-thick DCO$^+$ and DCN lines. The shaded regions correspond to the 1$\sigma$ uncertainty range.}
    \label{fig:taumax}
\end{figure}

\begin{figure}[t!]
    \centering
    \includegraphics[width=0.45\textwidth]{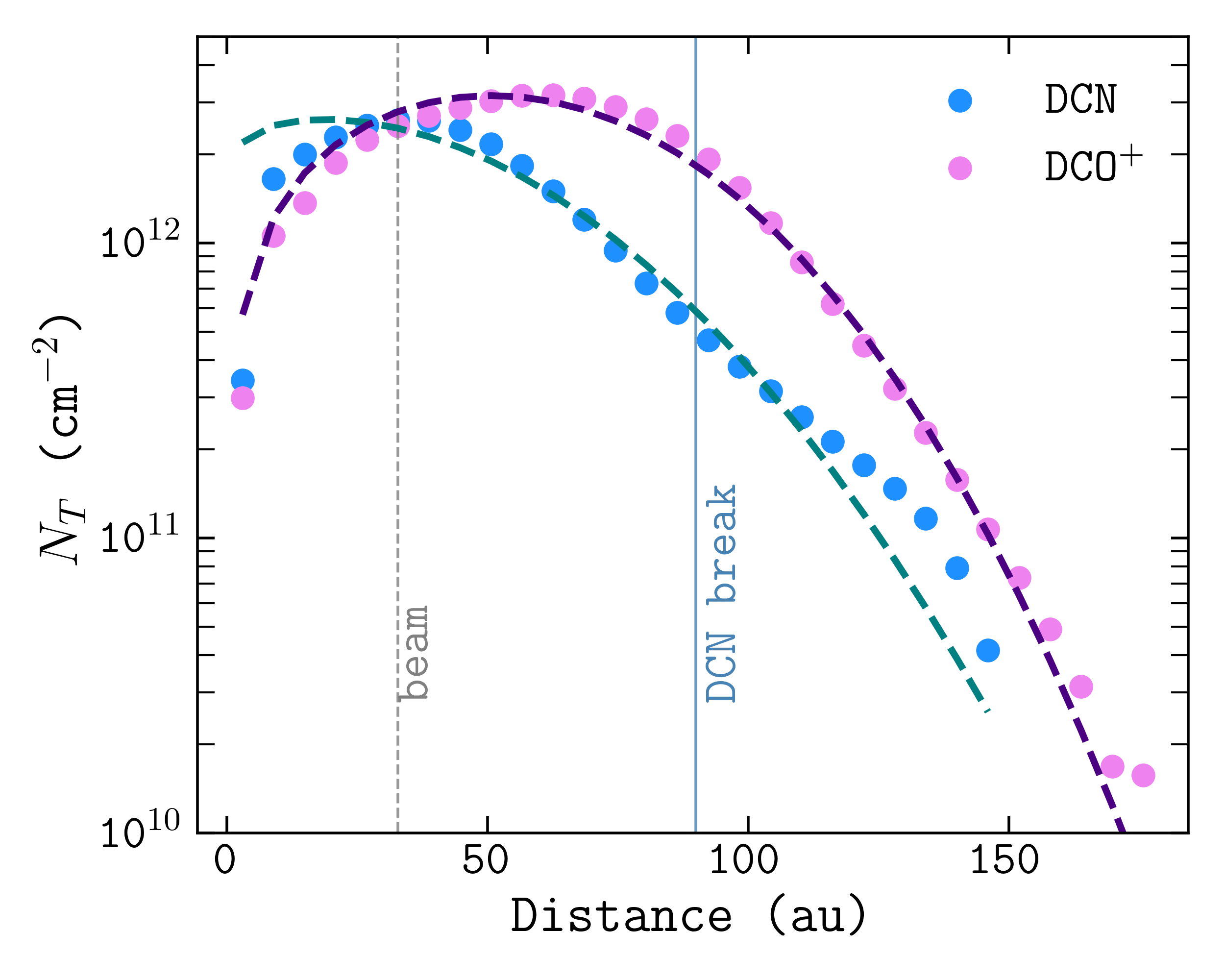}
    \caption{Radial column density profiles (scatter points) for each species and their best fit tapered power-law distribution (dashed lines).}
    \label{fig:powerlaw}
\end{figure}

We find that both the DCO$^+$ and DCN emission remain optically thin throughout the disk, such that we can obtain a precise estimate of the total column density, rather than a lower limit. Figure \ref{fig:taumax} shows the line-center optical depth for the most optically thick line of each species throughout the disk. We find a peak DCO$^+$ column density of $3.2\times10^{12}$ cm$^{-2}$ in the $59-65$ au radial bin, coincident with the peak integrated intensity, and a disk-averaged $N_{T}$ of $1.9\times10^{12}$ cm$^{-2}$ estimated inside the high S/N region enclosed by the dashed lines in Figure \ref{fig:DCO+_profiles}. The ring-like structure of the column density profile is less pronounced than that of the integrated intensity, with a roughly constant $N_{T}$ plateau spanning from $30-70$ au which only drops below 10$^{12}$ beyond 110 au. Similarly, we find a peak DCN column density of $2.6\times10^{12}$ cm$^{-2}$, closer in at the $30-36$ au radial bin, and a disk average of $9.8\times10^{11}$ cm$^{-2}$. Although less evident due to the logarithmic scale used, the DCN column density profile exhibits a break at around 90 au coincident to that exhibited by the intensity profiles in Figure \ref{fig:intprofiles}. This is evidenced more clearly if we model the column density profiles using an archetypal tapered power-law distribution of the form 
\begin{equation}
    N_T = N_0 \left( \frac{r}{r_c} \right)^{-\gamma} \exp \left[ -\left(\frac{r}{r_c}\right)^{2-\gamma} \right]
\end{equation}
\citep{Lynden-Bell_1974}. For DCO$^+$, we find the emission is very well described by a distribution with power-law index $\gamma = -0.70$ and characteristic radius $r_c = 84$ au, with $N_0 = 5.8\times10^{12}$ cm$^{-2}$. In the case of DCN, the best model has $\gamma = -0.13$, $r_c = 69$ au, and $N_0 = 3.3\times 10^{12}$ cm$^{-2}$. For comparison, the full gas disk is best described by $\gamma = 0.75$ and $r_c = 400$ au \cite{Calahan_2021}. As shown in Figure \ref{fig:powerlaw}, the model begins deviating considerably from the data beyond 90 au, where the emission break is observed. The model also over-predicts the DCN column density in the inner disk, but the significance of this discrepancy is unclear due to the limited spatial resolution. Together,  these deviations may indicate that there are contributions from multiple DCN formation pathways throughout the disk, as discussed in Section \ref{sec:discussion}. 


As indicated by the first panel of Figure \ref{fig:DCN_profiles}, we derive an excitation temperature of $\sim 25-30$K for DCN exterior to the inner emission depression, with a typical 1$\sigma$ lower limit of 19K and upper limit of 35K. There appears to be a bump at 90 au followed by a drop in the kinetic temperature below 19K at $r>r_{out}$, but the S/N is too low to properly interpret the latter feature. In the innermost, unresolved, disk regions, the temperature is observed to rapidly increase towards $\sim90$K, which could be simply a result of beam smearing. Similarly, DCO$^+$ exhibits a low kinetic temperature beyond $r>r_{in}$, with a roughly constant temperature of 27$\pm5$K, coincident with the CO freeze-out temperature derived by \citet{Zhang_2017}, from $\sim 20$ au out to 50 au, where it then decreases gradually down to 20K in the outer disk.  

For both DCN and DCO$^+$, we are unable to obtain tight constraints on the collider number density.  Instead, we find only a typical lower limit of $\sim 10^{7}$ cm$^{-3}$, above which the exact value of $n(H_2)$ has little to no impact on the derived column densities and excitation temperatures. We interpret this as a strong indication that LTE conditions dominate throughout the disk and, as mentioned before, we find no significant differences in our results when we carry out the fit assuming LTE. Furthermore, as seen in the fourth panel of Figures \ref{fig:DCN_profiles} and \ref{fig:DCO+_profiles}, the fits for the smoothing kernel width are consistent, with $\delta$ playing a significant role at small radii and swiftly falling below the typical thermal line width of 0.1 km/s at $r\gtrsim 50$ au, where the emission no longer needs to be substantially redistributed. At a comparable distance, the fits for $v_{sys}$ converge to a typical value of 2.85 km/s for both molecules.

\subsubsection{Estimating the Emitting Layer Height} \label{sec:zr}

We next attempt to approximate the emission layer height of DCN and DCO$^+$ by coupling the results of our excitation analysis with the 2D disk thermal structure model derived by \citet{Calahan_2021} as part of the TW Hya Rosetta Stone Project. The temperature model was calculated using \texttt{RAC2D} \citep{Du_2014} and accurately reproduces the emission profiles of two CO lines, two $^{13}$CO lines, three C$^{18}$O lines, and one HD line. To estimate the emission height in the disk, we use the data behind Figure 6 of \citet{Calahan_2021} (private communication), which corresponds to a grid of kinetic temperatures for radii between 0.1 and 200 au in 2 au increments, and heights between 0 and 100 au in 0.5 au increments. The model values are linearly interpolated to match the location where the DCN and DCO$^+$ temperatures were extracted, which allows us to solve for the height at each radial bin.  



\begin{figure}[t!]
    \centering
    \includegraphics[width=0.47\textwidth]{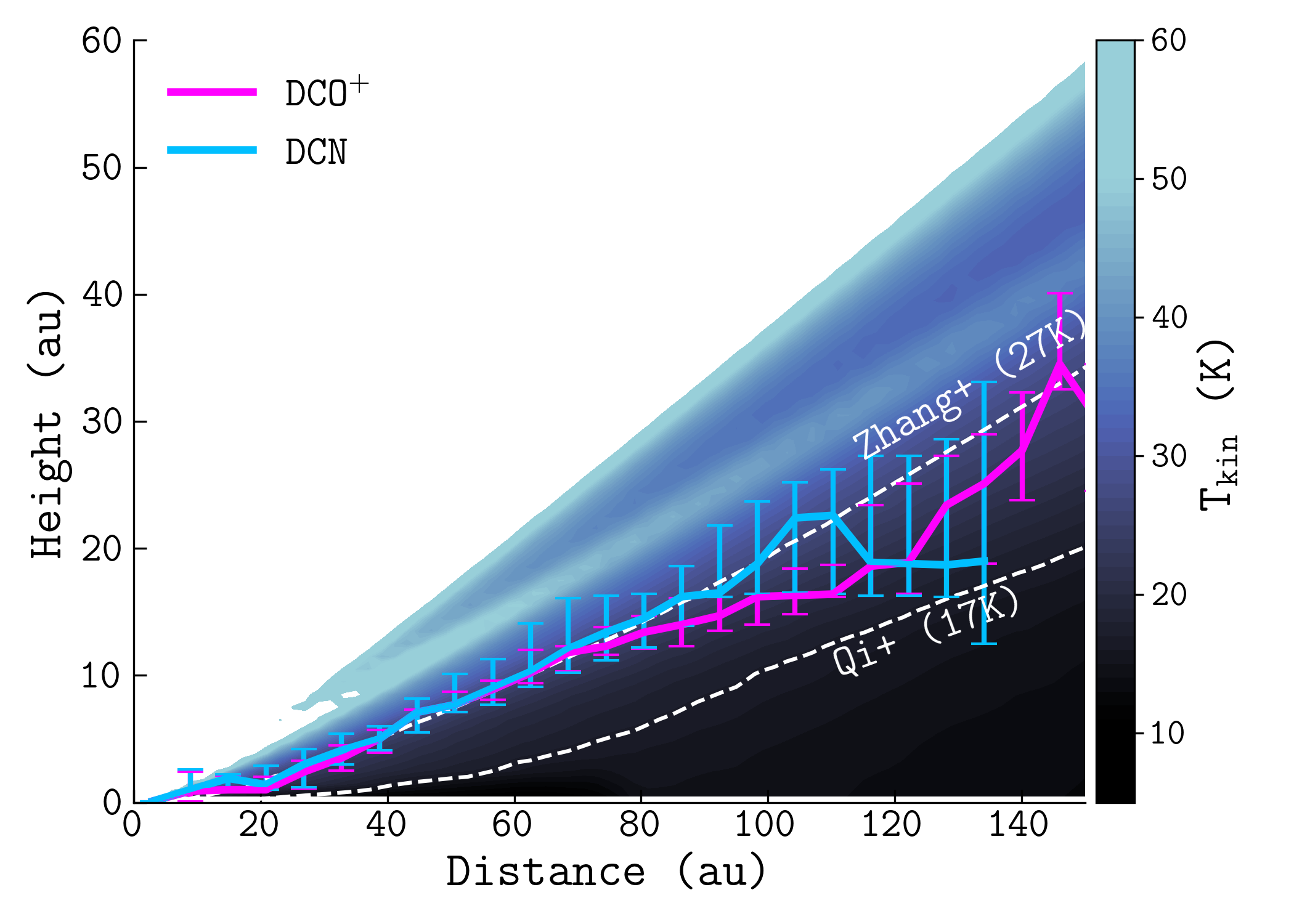}
    \caption{Estimated emission layer height of DCN and DCO$^{+}$ based on the estimated gas kinetic temperatures from \citet{Calahan_2021} (background color map), compared to the estimated CO snow-surface based on the freeze-out temperatures reported by \citet{Qi_2013} and \citet{Zhang_2017}. The error bars indicate the 1$\sigma$ confidence intervals.}
    \label{fig:z_r}
\end{figure}

Figure \ref{fig:z_r} shows the resulting emission height profiles for DCN and DCO$^+$ for the disk region where an emission height could be constrained, together with the thermal structure and the predicted CO snow-surface. The CO freeze-out temperature in TW Hya has been estimated to be between 17 and 27K based on N$_2$H$^+$ emission \citep{Qi_2013} and high-resolution CO isotopologue imaging \citep{Zhang_2017}, respectively. For both molecules, the emission layer height  is found to lie within both snow-surface limits throughout the disk, ranging from $z/r = 0.05$ in the inner disk to 0.25 in the outer disk. 

\section{Discussion}
\label{sec:discussion}

\subsection{Deuterium Fractionation in TW Hya}
\label{sec:deuteration}

We find evidence of active deuterium fractionation near the midplane of TW Hya, dominated by cold chemical reactions not unlike those observed in dark clouds and cores. Primarily, our excitation analysis reveals that DCN and DCO$^+$ are mainly emitting from a dense disk layer with a temperature between 20 and 30K, consistent with emission near the midplane. The similarity in the derived emitting gas kinetic temperatures indicates that the deuterated species share a cold formation pathway, as opposed to one dominated by warm hydrocarbons. This conclusion is reinforced by the fact that the derived excitation temperatures do not drop below the expected lower limit of 17K for the CO freeze-out temperature (albeit allowed to by our prior parameter distributions), which would indicate some emission originates from sub-thermally excited atmospheric layers. However, our 1D analysis cannot rule out the presence of a secondary emitting layer at higher disk elevations.

Similarly, the derived emitting layer height for both species is consistent with \textit{ab initio} predictions for a H$_2$D$^+$-driven disk deuterium fractionation \citep[e.g.][]{Mathews_2013}. Following Equation \ref{eq:colddco} and \ref{eq:colddcn}, the formation of DCN and DCO$^+$ requires abundant gas phase CO and H$_2$D$^+$, which can be problematic, since H$_2$D$^+$ forms directly from H$_3^+$ via 
$$
\mathtt{H_3^+ + HD \rightarrow H_2D^+ + H_2}
$$
and H$_3^+$ is readily destroyed by CO. The primary layer traced by the deuterated species, then, would be expected to be right at the interface of the CO freeze-out layer, where both reactants can coexist. As seen in Figure \ref{fig:z_r}, we find the emission heights of DCN and DCO$^+$ to agree and trace the predicted CO snow-surface remarkably well throughout the disk. The one exception occurs at $\sim95$ au, where DCN exhibits a 10K temperature bump not observed in DCO$^{+}$, which is coincident with the emission break seen in Figure \ref{fig:intprofiles}. While the uncertainty is large, it is likely this feature is real and related to increased UV penetration at the dip in scattered-light \citep{Debes_2013} and surface density perturbation traced by CS \citep{Teague_2017} observed at the same location.
 
Our results further offer a straightforward explanation for the inner emission depression exhibited by DCN and DCO$^+$: the H$_2$D$^+$-driven deuterium fractionation pathway simply shuts off in the inner disk. As illustrated by Figure \ref{fig:intprofiles}, the integrated intensity of both species drops quickly at distances smaller than 30 au, and the average inner ring radii fit to DCO$^+$ and DCN are very close to the most recent midplane CO snowline measurements of $17-23$au \citep{Schwarz_2016, Zhang_2017}. Furthermore, as discussed by Ö21, it is unlikely that the inner depression is an effect of dust opacity rather than a tracer of chemical processes, since optically thin species such as $^{13}$CO, C$^{18}$O, and $^{13}$C$^{18}$O \citep{Zhang_2017, Zhang_2019} exhibit centrally peaked emission profiles towards TW Hya.

Interestingly, the emission peak of DCN is located just interior to the CO snowline, which could indicate that some DCN is not necessarily bound to the abundance of H$_2$D$^+$ and there could be, in fact, lesser contributions from the warm hydrocarbon pathway. While the emission in this regime is unresolved, this scenario is not inconsistent with our results. First, there is abundant DCN down to $\sim 10$ au, and as shown by Figure \ref{fig:DCN_profiles}, the best fit temperature profiles become consistent with a temperature $\gtrsim40$K within $1\sigma$ at $<30$ au. Second, while the maximum DCO$^+$ abundance is found at 60 au, when the derived temperature drops to 20K, the peak DCN column density occurs closer in, at 33 au, when the temperature is still $>$25K. 

There are no multi-line observations of HCO$^+$ and HCN at the angular resolution necessary to carry out a similar excitation analysis as the one presented in this Paper. The existing observations of HCO$^+$ $J=3-2$, and $4-3$ \citep{Qi_2008, Cleeves15} show a similar ring morphology as that exhibited by DCO$^+$, consistent with a shared CO-mediated formation pathway. High-resolution observations of HCN $J=4-3$ \citep{Hily-Blant_2019}, on the other hand, exhibit a centrally-peaked distribution. The radial dependence of HCN deuteration has been observed in a few other disks \citep{Cataldi21}, and highlights the fact that deuterium fractionation is an \textit{in situ}, primarily temperature-regulated process.

In summary, exterior to the CO snowline, DCO$^+$ and DCN are predominantly co-located and forming through a cold pathway; in the innermost disk, both molecules are largely absent, indicating that a warm deuterium pathway is not active. At intermediate temperatures, however, DCO$^+$ is absent, but some DCN is still forming through a lukewarm, likely hydrocarbon-mediated pathway. Why the latter channel shuts down at higher temperatures is not entirely clear, however. On the one hand, it is possible that the propagation of deuteration from carbon chains onto DCN is possibly quenched by an elevated ortho-to-para-H$_2$ ratio and/or increased vertical mixing in the inner disk \citep{Willacy_2015, Aikawa_2018}. Alternatively, it could be that CH$_2$D$^+$ itself is being prevented from forming due to a lack of chemically active gas-phase carbon in the inner disk \citep{Bergin_2016} or a lack of ionization in the inner disk. The latter has been observed in IM Lup \citep{Seifert_2021}, and would further explain why we find no evidence of a warm DCO$^+$ reservoir in the inner disk as suggested by \citet{Favre_2015}.

\subsubsection{Discrepancy with Previous Results}
\label{sec:oberg}

The average kinetic temperature derived for DCO$^+$ in TW Hya is smaller by $\sim15$K from that recently reported by Ö21, who used the same $J=3-2$ and $4-3$ line data. Although small, this divergence has critical implications for the inferred deuterium fractionation mechanism and its potential impact on forming planets, since the warm CH$_2$D$^+$ pathway is expected to dominate at 40K \citep{Wootten_1987, Favre_2015}. 

We explored the possible sources of this discrepancy by focusing on three differences between the two studies: different temperature and column density retrieval codes, a lack of non-Gaussianity (JvM) corrections in Ö21, and the new DCO$^+$ 2--1 data used for this study. Using our retrieval framework on the Ö21 data, we recovered their elevated DCO$^+$ temperature and therefore rule out this as a source of the discrepancy. Next, we removed the JvM correction from our data and reran the retrievals, but found negligible differences between the excitation temperatures. Finally, we carried out the retrievals again using only the archival 4--3, 3--2, and 2--1 observations, which resulted in an increase of $\sim10$K in the derived excitation temperature. This significant discrepancy highlights the critical importance of high-fidelity 2--1 data for these kind of excitation analyses, even if costly in terms of ALMA time.

The minor temperature difference that remained was likely due to different moment-zero map generation techniques between the two studies, i.e. multiple Gaussian fits for each hyperfine line component in this work as opposed to direct integration of the image cubes. By combining the $J=4-3$ and $J=3--2$ integrated intensity maps from the Ö21 study with the new $2-1$ observations, we recovered an additional $3-5$K discrepancy throughout the disk.

\subsection{Comparison with Other Disks}
\label{sec:otherdisks}

The abundance and excitation conditions of DCN and DCO$^+$ have only been studied towards a few other protoplanetary disks. Most recently, \citet{Cataldi21} observed DCN $J=3-2$ towards IM Lup, GM Aur, AS 209, HD 163296, and MWC 480 at high angular resolution (0.3"), and reported column densities between 10$^{11}$ and 10$^{13}$ cm$^{-2}$ for all disks, similar to that found in this work. From the DCN/HCN ratio, \citet{Cataldi21} found evidence that both cold and warm fractionation pathways must be active in these disks, with typical temperatures between 20 and 50K, which is consistent with the warmer temperatures we derive for DCN near the CO snowline. The DCN morphologies observed by \citet{Cataldi21}, however, are different and seem to have a strong dependence on the individual disk properties. Similarly, \citet{Salinas_2017} concluded that both fractionation pathways contribute towards the formation of DCN in HD 163296, reported an average column density of $2.9\times10^{11}$ cm$^{-2}$, comparable to that found in TW Hya. 

In the case of DCO$^+$, previous observations point towards a universal, cold, H$_2$D$^+$-dominated formation pathway in disks. In HD 163296, \citet{Mathews_2013}, \citet{Teague_2015}, and \citet{Flaherty_2017} also found DCO$^+$ to be located just exterior the CO snow-surface, and a similar plateau-like DCO$^+$ column density profile with an average value of $\sim10^{12}$ cm$^{-2}$, while \citep{Salinas_2017} argue that some contributions from the CH$_2$D$^+$ pathway must be active to explain the multi-ringed distribution of DCO$^+$ in HD 163296. \citet{Carney_2018}, likewise, found that most of the DCO$^+$ emission in HD 169142 originates in the cold ($<25$K) midplane with a column density of $3.7\times10^{11}$ cm$^{-2}$, albeit with significant contributions of the warm pathway in the inner disk. 

\subsection{Column Density Ratios}

One potentially intriguing result of our analysis is the relatively low DCO$^+$/DCN column density ratio in the inner disk. Note, however, that the emission interior to 30 au is unresolved, and thus any inferences regarding the molecular ratios in this regime must be confirmed by higher resolution observations. At 10 au, the derived DCO$^+$/DCN ratio is $0.7$, it remains $<$1 inside the CO snowline (20--30 au), and only begins to increase at $40$ au. The ratio then increases to $\sim6$ at 90au and drops again to $<2$ in the outermost regions. While the observed DCO$^+$ abundance of $\sim10^{12}$cm$^{-2}$ in TW Hya is consistent with most theoretical estimates and that of other disks, the abundance ratio of DCO$^+$ to DCN in TW Hya is approximately an order of magnitude lower than predicted by most models \citep{Aikawa_2002, Favre_2015}, which is also observed in HD 163296 \citep{Salinas_2017}. Only the models of \citet{Willacy_2007}---which account for efficient ice photodesorption---predict a DCO$^+$/DCN of order unity. We surmise the over-abundance of DCN observed in TW Hya relative to models is thus likely due to poorly understood grain surface chemical processes. Alternatively, it is possible the excess DCN is caused by an elevated C/O ratio, which may enhance the contributions from the CH$_2$D$^+$-dominated formation pathway. We explore this hypothesis below. 

We compare the observed DCO$^+$/DCN ratio with the modeling results of \citet{Aikawa_2018}. We choose these models since they were conveniently calculated for two cases: one for a C/O $<$1 based on the initial abundances of \citet{Furuya_2016}, and a CO- and H$_2$O-depleted case with an elevated C/O of 1.43. For complete details regarding the simulations, we direct the reader to the original paper \citep{Aikawa_2018}. In brief, the models were calculated for both C/O ratios at 1, 3, and 9.3$\times10^{5}$ years, for a T Tauri star of $M_{*} = 0.5M_{\odot}$, with a surface temperature of 4000K, and a UV and X-ray spectrum based on that of TW Hya, accounting for both vertical mixing and a varying ortho-to-para-H$_2$ ratio to best reproduce observations of DCO$^+$ and DCN. We emphasize, however, that the models fail to reproduce the observed abundance of HCN at small radii ($<$10au) \citep{Bergner_2021}, so this region is not included in the comparison. 

\label{sec:ratio}
\begin{figure}[ht!]
    \centering
    \includegraphics[width=0.47\textwidth]{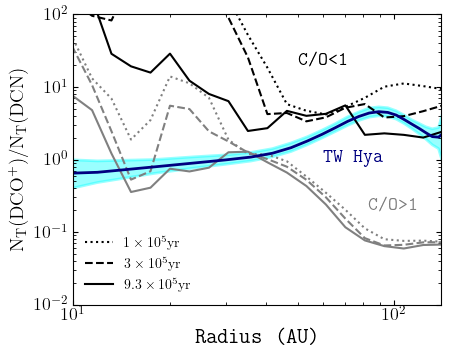}
    \caption{Column density ratio of DCO$^{+}$ to DCN. Shaded area in blue corresponds to the 1 $\sigma$ confidence region. The models from \citep{Aikawa_2018} are shown in black (for C/O$<1$) and grey (C/O $>1$).}
    \label{fig:density_raios}
\end{figure}

Figure \ref{fig:density_raios} shows the TW Hya observations and the modeled ratios for each case, from 10 to 110 au. Despite not being tuned to match the conditions of TW Hya, the models match the observed DCO$^+$/DCN remarkably well, but no single C/O can match the full radial DCO$^+$/DCN profile. Instead, we observe the DCO$^+$/DCN ratio in TW Hya to closely resemble the C/O $>$1 case out to 40 au, after which the observations appear to be very well described by a C/O $<$1. In both cases, the observations are best described by the 9.3$\times10^{5}$ year old models, which is not surprising for the old TW Hya disk.

Note that the simulations presented above were in no way calibrated to match the temperature and density structure of TW Hya. As such, the agreement between the models and our observational results can be best interpreted as an indication of a substantial gas-phase C/O ratio enhancement in the innermost 40 au compared to the outer disk, rather than evidence for a particular C/O value. It is possible this C/O enhancement in the inner disk traces the release of carbon into the gas phase due to increased release of active carbon from grains and destruction of large carbon chains by UV radiation, which would then be incorporated into CH$_2$D$^+$ and hence DCN. If true, then the DCN emission depression in the innermost regions may in fact be best explained by quenching due to a non-standard ortho-to-para-H2 ratio, as discussed before. 

Recent observations of C$_2$H \citep{Bergin_2016} and c-C$_3$H$_2$ \citep{Cleeves_2021} in TW Hya suggest that the upper molecular layers and the disk atmosphere, especially past 30 au, is rich in hydrocarbons and can only be described by models with C/O $>1$. To reconcile these results with our model comparison, one possibility is that, as mentioned above, we are simply tracing a relative gradient in the gas phase C/O, rather than a true C/O$<1$ in the outer disk. Alternatively, it is possible that the midplane C/O is indeed significantly lower than that of the upper molecular layers, which would further affirm our main conclusion that both DCN and DCO$^+$ originate in the former region. The face-on geometry of TW Hya makes it intrinsically difficult to study elemental ratio tracers near the midplane, and further modeling work as well as observations of more inclined disks will be necessary to confirm whether DCN and DCO$^+$ can in fact serve as tracers of radial and vertical C/O ratio gradients.

\section{Summary}
\label{sec:conclusions}
\begin{enumerate}
    \item We present new ALMA Band 4 observations of DCO${^+}$ and DCN $J=2-1$, augmented with archival data of the $J=3-2$ and $J=4-3$ transitions. All transitions of both molecules exhibit ring-like emission patterns: DCN shows a narrow ring feature peaking at 40 au and an emission break at 90 au, likely due to increased UV penetration, while DCO$^+$ shows an extended plateau that peaks at 60 au and extends further out in the disk. 
    
    \item We carry out a detailed 1D non-LTE analysis and model the azimuthally averaged and deprojected spectra using \texttt{RADEX} and Bayesian retrieval framework to estimate the column density and excitation conditions traced by each molecule. We derive similar column densities of $\sim 10^{12}$ cm$^{-2}$ for both species and excitation temperatures in the range of 20-30K. We are only able to provide a lower limit to the H$_2$ gas density of $\sim10^6$ cm$^{-3}$.
    
    \item Our results imply that both DCN and DCO$^+$ emit primarily from a region just above the CO snow-surface, and their formation is dominated by the cold H$_2$D$^+$ pathway. There is evidence that some DCN may be formed via the CH$_2$D$^+$ pathway in the lukewarm disk regions, but it likely is quenched by an elevated ortho-to-para H$_2$ ratio in the inner disk.
    
    \item The discrepancy between our results and previous findings by Ö21 are best explained by the increased sensitivity of our $J=2-1$ observations and, to a lesser extent, by the full treatment of the hyperfine line structure exhibited by both molecules in the integrated intensity map generation. 
    
    \item We explore the possibility that the elevated DCN/DCO$^+$ ratio observed in TW Hya is due to a super-solar C/O ratio. A comparison with modeling results by \citet{Aikawa_2018} suggests a relative enhancement in the C/O ratio in the inner disk, which could be caused by carbon chain destruction. It remains unclear whether the C/O ratio in the outer disk midplane is truly $<1$, and thus if it is significantly lower than that of the disk upper molecular layers.
    
    \item Observations of DCN and DCO$^+$ along with their non-deuterated analogs in a more diverse sample of disks will be essential to confirm if deuterium fractionation is indeed dominated by low-temperature chemical processes, and thus to fully constrain the main deuteration mechanisms in disks.

\end{enumerate}

The authors thank the anonymous referee for valuable comments that enhanced the content of this work. C.E.R.M. would like to thank Sean Andrews for fruitful conversations on emission line modelling, and Jenny Calahan for sharing the \texttt{RAC2D} thermal structure grid. 

This paper makes use of the following ALMA data: ADS / JAO.ALMA\# 2016.1.00311.S, ADS / JAO.ALMA\# 2016.1.00440.S, ADS / JAO.ALMA\# 2017.1.00769.S, and ADS / JAO.ALMA\# 2019.1.01153.S. ALMA is a partnership of ESO (representing its member states), NSF (USA) and NINS (Japan), together with NRC (Canada), MOST and ASIAA (Taiwan), and KASI (Republic of Korea), in cooperation with the Republic of Chile. The Joint ALMA Observatory is operated by ESO, AUI/NRAO, and NAOJ. This work is supported by the National Radio Astronomy Observatory (NRAO). NRAO is a facility of the National Science Foundation operated under cooperative agreement by Associated Universities, Inc. K.I.Ö. acknowledges support from the Simons Foundation (SCOL \#321183) and an NSF AAG Grant (\#1907653). C.J.L. acknowledges funding from the National Science Foundation Graduate Research Fellowship under Grant No. DGE1745303. J.B.B. acknowledges support from NASA through the NASA Hubble Fellowship grant \#HST-HF2-51429.001-A awarded by the Space Telescope Science Institute, which is operated by the Association of Universities for Research in Astronomy, Incorporated, under NASA contract NAS5-26555. V.G. acknowledges support from FONDECYT Regular 1221352, ANID BASAL projects ACE210002 and FB210003, and ANID, -- Millennium Science Initiative Program -- NCN19\_171. 

\facility{ALMA}

\software{Numpy \citep{numpy}, Astropy \citep{astropy:2022}, keplerian\_mask.py \citep{keplerian_mask}, RADEX \citep{vandertak_2007}, CASA \citep{McMullin_2007}, dynesty \citep{Speagle_2020}}.

\appendix

\section{Line Modeling} \label{sec:modelapx}

Following the approach presented by \cite{Teague_2018} and \cite{Bergner_2021}, all emission lines are assumed to be well-described by a Gaussian profile that accounts for saturation at the line core, such that the total optical depth is
\begin{equation}
    \tau_v = \sum_i \tau_{i,0} \exp\left(-\frac{1}{2}\frac{(v-v_i-v_{sys})^2}{\sigma^2}\right),
     \label{eq:tau}
\end{equation}
where each subscript $i$ corresponds to a hyperfine line component, $v$ is the velocity, $v_{sys}$ the systemic velocity, $\tau_{i,0}$ the line-center optical depth, and $\sigma$ the Gaussian line-width. Then, taking advantage of the low inclination of the TW Hya disk, a plane-parallel geometrically averaged escape probability is adopted,
\begin{equation}
    \beta = \frac{1-e^{-3\tau}}{3\tau},
\end{equation}
and the azimuthally averaged spectra is modeled as
\begin{equation}
    T_{B} = (J_v(T_{ex}) - J_v(T_{bg}))\times(1-e^{-\tau_v}),
\end{equation}
where $T_{bg}$ is the background temperature, assumed to be equal to $T_{\text{CMB}}=2.73$K, $T_{ex}$ the molecular excitation temperature, and $$ J_v(T) = \frac{h \nu /k}{\exp(h \nu/kT)-1} $$ is the Planck function. 

The line widths, in principle, include thermal and non-thermal broadening components, such that
\begin{equation}
\sigma = \sqrt{\frac{kT_{kin}}{\mu m_{H}} + v_{turb}^2},
\label{eq:sigma}
\end{equation}
where $\mu$ is the mean molecular weight of each species, $v_{turb}$ the line-of-sight velocity dispersion, and $T_{kin}$ the gas kinetic temperature. Multiple studies, however, have provided upper limits for the effects of turbulent broadening in disks, generally finding that turbulent broadening effects are minor and, in the case of TW Hya,  $\mathcal{M}\lesssim0.4$ \citep{Flaherty_2018, Teague_2016, Teague_2018}, where the Mach number is $\mathcal{M} = v_{turb}/c_s$ and $c_s$ is the local sound speed. Thus, we decide not to consider the effects of turbulent line broadening in our modeling. Nonetheless, we observe an artificial broadening of the lines produced primarily by the beam-smearing of multiple velocity components onto individual pixels, and aggravated by the imperfect stacking of the lines, the Hanning smoothing of the data, and possible contributions from the front and back sides of the disk \citep{Bergner_2021}. In the case of the Keplerian shifting, we emphasize that the imperfections lie not in the shifting of the lines, but rather in the assumption of the background velocity structure, which may not be purely Keplerian \citep{Teague_19}. These effects make it impossible to model the spectra in the innermost $\sim 30$ au assuming simply a thermal line width, and drive the retrieved kinetic temperatures to unreasonably high values ($> 100$ K) throughout the disk, since the line widths provide a strong constraint on the kinetic temperature in this modeling framework.

Following \citet{Cataldi21}, we account for the non-physical broadening of the lines by convolving the model spectra with a Gaussian kernel of width $\delta$--which we treat as a free parameter--to redistribute the integrated intensity and match the observed line widths. The spectra is modeled at $100\times$ the sampling rate of the observations and binned to the data resolution prior to carrying out the convolution with $\delta$.\footnote{Based on the results of \citet{Teague_2018}, we do not attempt to model any spectral correlations in the noise.} By redistributing the emission we lose degrees of freedom and, primarily, information regarding the local gas density. Nonetheless, we decide to treat $\log_{10}n(H_2)$ as a free parameter in order to avoid introducing any model dependencies into our results.

To account for potential non-LTE effects, the 1D radiative transfer code \texttt{RADEX} \citep{vandertak_2007}\footnote{Specifically, we use the Python wrapper \texttt{pyradex}: https://github.com/keflavich/pyradex} was coupled with the dynamic nested sampling Python package \texttt{dynesty} \citep{Speagle_2020} and used as a forward modeling tool. We calculate excitation temperatures, model spectra, and estimate Bayesian posterior distributions for $T_{kin}$, $\log_{10}N_T$, $\log_{10}n(H_2)$, $v_{sys}$, and $\delta$. In the absence of hyperfine collisional rates for DCO$^{+}$ and DCN, we create our own molecular data files for \texttt{RADEX} using collisional data for HCO$^{+}$--H$_2$ \citep{Flower_1999} and HCN--H$_2$ \citep{Vera14}, respectively, and spectroscopic information from the CDMS \citep{muller_05} and JPL \citep{Pickett_1998} databases for DCN and DCO$^{+}$, respectively. While the isotopic substitution fundamentally changes the molecular symmetry, and hence the potential surface of interaction, the effects on the collisional properties for DCO$^{+}$ and DCN  are expected to be small \citep[e.g.][]{Buffa_2012, Dennis_20} relative to the intrinsic uncertainty of the collisional rates (factor of $\sim2$). 

For a given $T_{kin}$, $\log_{10}N_T$, $\log_{10}n({H_2})$, and $\beta$, excitation temperatures are calculated for each $J-$transition with \texttt{RADEX}, and the line-center optical depth of each satellite line is estimated as 
\begin{equation}
    \tau_{i,0} = \frac{N_T}{Q(T_{ex})} e^{-\frac{E_u}{T_{ex}}} \frac{g_u A_{ul} c^3}{8 \pi \nu^3 \sqrt{2\pi}\sigma} \left(e^{\frac{h \nu}{k_B T_{ex}}} - 1\right),
    \label{eq:taulte}
\end{equation}
where $T_{ex}$ is assumed to be the same for all $F-$hyperfine components. Here, $g_u$ is the upper energy level degeneracy, $A_{ul}$ the Einstein coefficient of each transition, ${E_u}$ the upper level energy, and $Q$ is the rotational partition function,
$$
Q(T) = \sum_i g_{u,i} \exp(-E_{u,i}/T),
$$
obtained from the same catalogues mentioned above and linearly interpolated to match the modeled kinetic temperature. The line coefficients used for each transition, along with reference values for each partition function, are presented in Appendix \ref{sec:apb}. Finally, note that $\beta$ is a parameter that is only used internally in \texttt{RADEX} to solve the radiative transfer equation.

Equipped with Equations (\ref{eq:tau}) $-$ (\ref{eq:taulte}) and \texttt{RADEX}, a likelihood function is constructed and the parameter space is sampled with \texttt{dynesty} using uniform prior distributions informed by previous observations of the TW Hya disk:

\begin{equation} \label{priors}
\begin{split}
\log_{10}N_{T} \mathrm{ (cm^{-2})} & = \mathcal{U}(10,14), \\
 T_{kin} \mathrm{ (K)} & = \mathcal{U}(10,150), \\
 \log_{10}n(H_2) \mathrm{ (cm^{-3})} & = \mathcal{U}(4,11), \\
 v_{sys} \mathrm{ (km/s)} & = \mathcal{U}(1,5), \\
\delta \mathrm{ (km/s)} & = \mathcal{U}(0, 3),
\end{split}
\end{equation}
where $\mathcal{U}(x,y)$ denotes a uniform prior distribution with lower and upper limits $x$ and $y$. All the transitions in each radial bin are fit simultaneously, and runs are terminated once the estimated contribution of the remaining prior volume to the total evidence falls below 0.1. 

\section{Spectral Line Information} \label{sec:apb}
Table 3 shows the full spectroscopic line data for DCN and DCO$^+$, and Table 4 shows representative values of the  DCN and DCO$^+$ partition functions calculated in \texttt{RADEX}. 

\startlongtable
\begin{deluxetable*}{llcccc}
\tabletypesize{\footnotesize}
\tablewidth{0pt}

 \tablecaption{Molecular Parameters \label{tab:moline}}

 \tablehead{
 \colhead{Species} & \colhead{Transition} & \colhead{Frequency (GHz)}  & \colhead{$E_u$ (K)$^{\dagger}$} & \colhead{log$_{10}$A$_{i,j}$ (s$^{-1}$) $^{\ddagger}$} & \colhead{$g_u$ $^{\ast}$} 
 }
 \startdata 
 DCN & $J=2-1 , F=2-2$ & 144.82657570 & 10.42589 & -4.49995 & 5  \\
 DCN & $J=2-1 , F=1-0$ & 144.82682160 & 10.42604 & -4.15320 & 3  \\
 DCN & $J=2-1 , F=2-1$ & 144.82800110 & 10.42595 & -4.02284 & 5  \\
 DCN & $J=2-1 , F=3-2$ & 144.82811060 & 10.42596 & -3.89787 & 7  \\
 DCN & $J=2-1 , F=1-2$ & 144.82890720 & 10.42600 & -5.45419 & 3  \\
 DCN & $J=2-1 , F=1-1$ & 144.83033260 & 10.42607 & -4.27809 & 3  \\
 DCN & $J=3-2 , F=3-3$ & 217.23699900 & 20.85157 & -4.29392 & 7  \\
 DCN & $J=3-2 , F=2-1$ & 217.23830000 & 20.85177 & -3.41539 & 5  \\
 DCN & $J=3-2 , F=3-2$ & 217.23855500 & 20.85164 & -3.39082 & 7  \\
 DCN & $J=3-2 , F=4-3$ & 217.23861200 & 20.85164 & -3.33966 & 9  \\
 DCN & $J=3-2 , F=2-3$ & 217.23907900 & 20.85167 & -5.69179 & 5  \\
 DCN & $J=3-2 , F=2-2$ & 217.24062200 & 20.85174 & -4.14779 & 5  \\
 DCN & $J=4-3 , F=4-4$ & 289.64331300 & 34.75225 & -4.15325  & 9  \\
 DCN & $J=4-3 , F=3-2$ & 289.64480300 & 34.75247 & -2.98610  & 7  \\
 DCN & $J=4-3 , F=4-3$ & 289.64492100 & 34.75233 & -2.97714  & 9  \\
 DCN & $J=4-3 , F=5-4$ & 289.64495700 & 34.75233 & -2.94909  & 11  \\
 DCN & $J=4-3 , F=3-4$ & 289.64529750 & 34.75235 & -5.84340  & 7 \\
 DCN & $J=4-3 , F=3-3$ & 289.64689700 & 34.75242 & -4.04400  & 7  \\
 DCO$^+$ & $J=2-1 , F=1-1$ & 144.07721440 & 10.37194	 & -4.55000  & 3 \\
 DCO$^+$ & $J=2-1 , F=1-2$ & 144.07726190 & 10.37194 & -5.72610  & 3  \\
 DCO$^+$ & $J=2-1 , F=3-2$ & 144.07728030 & 10.37194 & -4.16988  & 7  \\
 DCO$^+$ & $J=2-1 , F=2-1$ & 144.07728510 & 10.37194 & -4.29475  & 5  \\
 DCO$^+$ & $J=2-1 , F=1-0$ & 144.07732360 & 10.37194 & -4.42510  &  3 \\
 DCO$^+$ & $J=2-1 , F=2-2$ & 144.07733260 & 10.37194 &  -4.77185 &  5 \\
  DCO$^+$ & $J=3-2 , F=2-2$ & 216.11251820 & 20.74365 & -4.41977  & 5  \\
 DCO$^+$ & $J=3-2 , F=2-3$ & 216.11257050 & 20.74365 & -5.96376  & 5  \\
 DCO$^+$ & $J=3-2 , F=4-3$ & 216.11257660 & 20.74365 & -3.61164  & 9  \\
 DCO$^+$ & $J=3-2 , F=3-2$ & 216.11257990 & 20.74365 & -3.66279  & 7  \\
 DCO$^+$ & $J=3-2 , F=2-1$ & 216.11258890 & 20.74365 & -3.68736  & 5  \\
 DCO$^+$ & $J=3-2 , F=3-3$ & 216.11263220 & 20.74365 & -4.56589  & 7  \\
 DCO$^+$ & $J=4-3 , F=3-3$ & 288.14380070 & 34.57223 & -4.31596  & 7  \\
 DCO$^+$ & $J=4-3 , F=5-4$ & 288.14385500 & 34.57224 & -3.22106  & 11  \\
 DCO$^+$ & $J=4-3 , F=3-4$ & 288.14385640 & 34.57224 & -6.11536  & 7  \\
 DCO$^+$ & $J=4-3 , F=4-3$ & 288.14385770 & 34.57224 & -3.24901  & 9  \\
 DCO$^+$ & $J=4-3 , F=3-2$ & 288.14386240 & 34.57224 & -3.25796  & 7  \\
 DCO$^+$ & $J=4-3 , F=4-4$ & 288.14391330 & 34.57224 & -4.42511  & 9  \\
 \enddata

 \tablecomments{$\dagger$Upper level energy. $\ddagger$Einstein coefficient. $\ast$Upper state degeneracy. Parameters for DCN were obtained from the CDMS molecular catalogue \citep{muller_05, Mollmann_2002, Bruken_2004}. Those for DCO$^+$ were obtained from the JPL catalogue \citep{Pickett_1998, Caselli_2005, Lattanzi_2007}.}
\end{deluxetable*}

\begin{deluxetable}{ccc}[ht!]
\tabletypesize{\footnotesize}
\tablewidth{0pt}
 \tablecaption{Partition Functions  \label{tab:qpart}}
 \tablehead{
 \colhead{Temperature (K)} & \colhead{$Q_{\mathrm{DCN}}$} & \colhead{$Q_{\mathrm{DCO^+}}$}}
 \startdata 
     9.3 & 14.22 & 4.77 \\
     18.8 & 30.39 & 10.19 \\
     37.5 & 62.75 & 21.03 \\
     75.0 & 127.50 & 42.73 \\
     150.0 & 256.93 & 86.13 \\
     225.0 & 384.50 & 129.46 \\
 \enddata
 \tablecomments{Representative values of the DCN and DCO$^+$ partition functions as calculated by \texttt{RADEX}}
\end{deluxetable}

\section{Spectral Fits}
\begin{figure*}[ht!]
    \centering
    \includegraphics[width=0.99\textwidth]{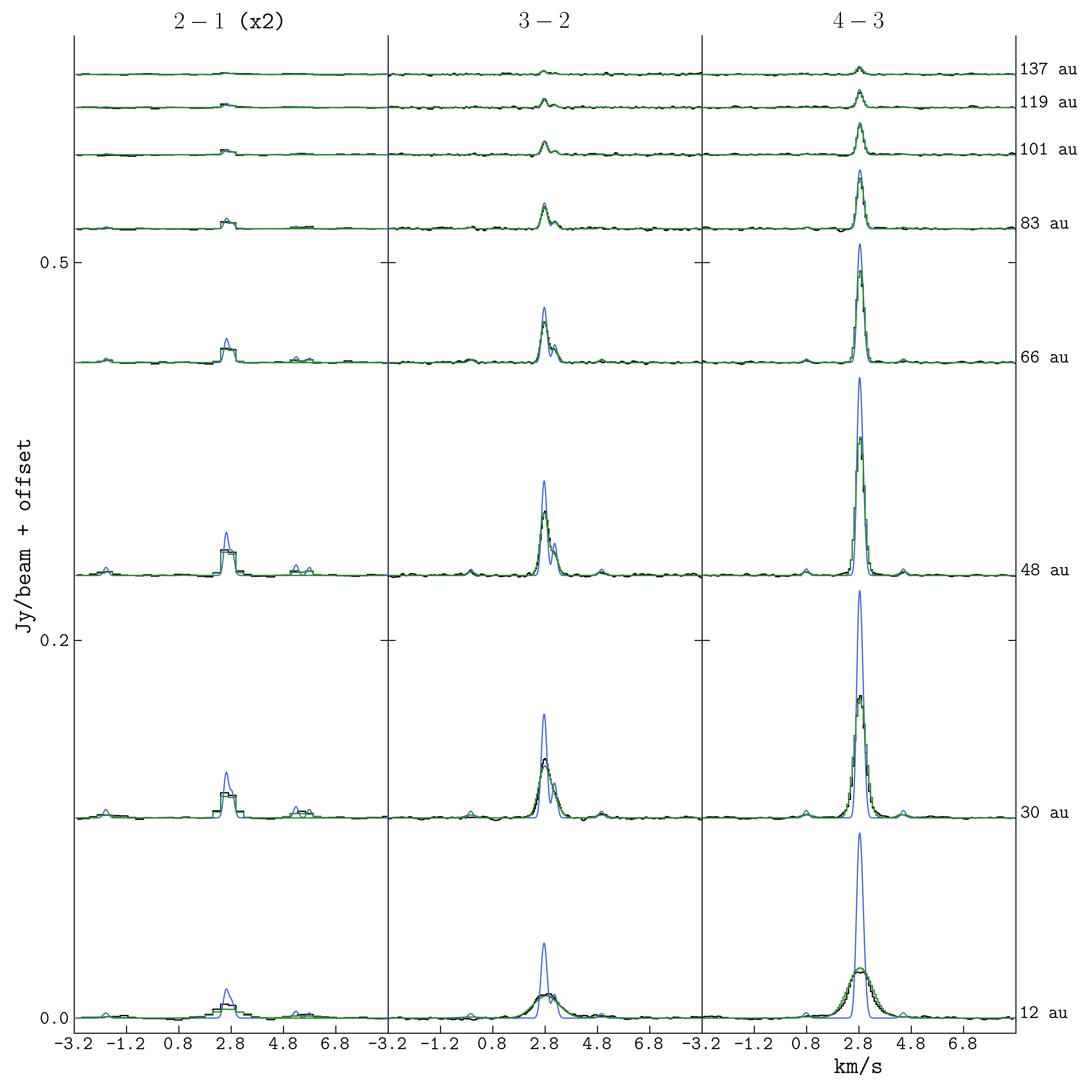}
    \caption{Gallery of azimuthally stacked and deprojected spectra of DCN $J=$2--1 (left), 3--2 (middle), and 4--3 (right). The blue profiles indicate the best-fit \texttt{RADEX} model, the green profiles show the best-fit model after convolution and down-sampling to match the observational resolution while conserving the total flux, and the black profiles correspond to the observations. Note that the $J=2-1$ spectra have been scaled by a factor of 2 to better display the profiles. The radii in au at which the spectra were extracted are indicated at the right.}
    \label{fig:dcngal}
\end{figure*}

\begin{figure*}[ht!]
    \centering
    \includegraphics[width=0.99\textwidth]{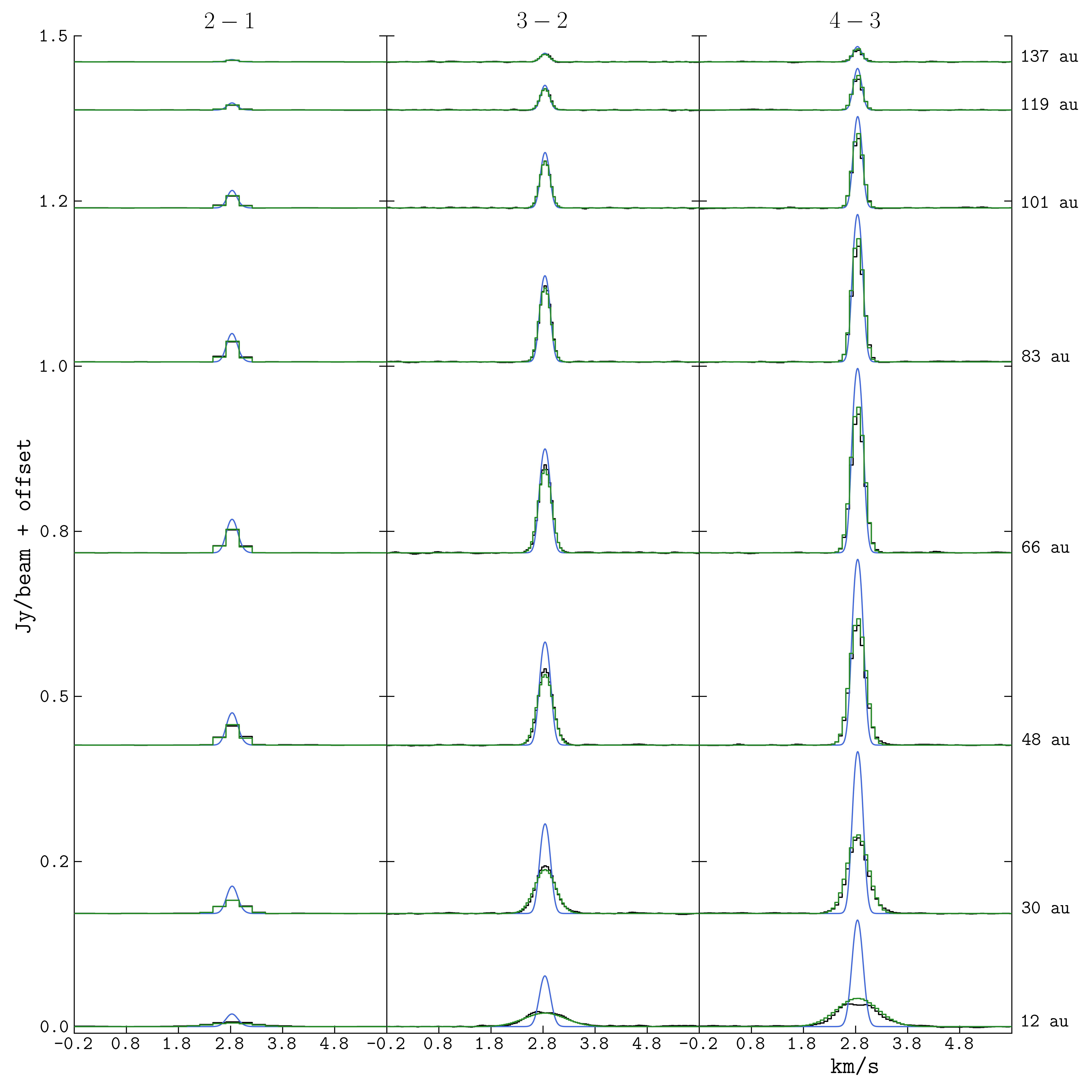}
    \caption{Same as Figure \ref{fig:dcngal}, for DCO$^+$}
    \label{fig:dcopgal}
\end{figure*}

Figures \ref{fig:dcngal} and \ref{fig:dcopgal} show a gallery of azimuthally stacked and deprojected spectra of DCN and DCO$^+$.

\bibliography{deuteration}{}
\bibliographystyle{aasjournal}



\end{document}